\documentclass[epj,final]{svjour}
\input epsf
\input psfig
\begin{document}
\title{Matter-induced $\omega \to \pi \pi$ decay%
\thanks{Research supported by PRAXIS grants XXI/BCC/429/94 and
        PRAXIS/P/FIS/12247/1998, and by
        the Polish State Committee for
        Scientific Research grant 2P03B-080-12.}}
\author{Wojciech Broniowski\inst{1} \and Wojciech Florkowski\inst{1}
\and Brigitte Hiller\inst{2}%
}                     
%
%
\institute{H. Niewodnicza\'nski Institute of Nuclear Physics,
         PL-31342 Krak\'ow, Poland,
         \email{broniows@solaris.ifj.edu.pl, florkows@solaris.ifj.edu.pl}\and
         Center for Theoretical Physics,
   University of Coimbra, P-3000 Coimbra, Portugal,
   \email{brigitte@malaposta.fis.uc.pt}}
\date{Received: date / Revised version: date}
%
\abstract{
We calculate the width for the $\omega \to \pi \pi $ decay in nuclear
matter in a hadronic model including mesons, nucleons and $\Delta$
isobars. We
find a substantial width of the longitudinally polarized $\omega$ modes,
reaching  $\sim 100$MeV for mesons moving suitably fast with respect to the
nuclear medium.
} 
\maketitle
The dilepton measurements in the CERES \cite{ceres} and HELIOS \cite{helios}
experiments have indicated that the masses and/or widths of light vector
mesons undergo large modifications in nuclear matter. Clearly, since the
mesons interact strongly with the medium, this fact is not at all
surprising. Indeed, in-medium modifications of hadron properties are
predicted in a variety of theoretical calculations \cite
{brscale,celenza,hatlee,jean,cassing,li,hatsuda,rapp,%
pirner,klingl,leupold,eletsky,friman2,bratko} (for a
recent review see \cite{hadrons,tsukuba}). An interesting factor brought in
by the presence of the medium is that processes which are forbidden in the
vacuum by symmetry principles are now made possible. The constraints of
Lorentz-invariance, $G$-parity, or isospin invariance in isospin-asymmetric
media \cite{dutt,rhoom}, are no longer effective. An example of such an
``exotic'' phenomenon which becomes possible and significant in the presence
of nuclear matter is the decay of $\omega \to \pi \pi $. This process is,
apart for small isospin-violation effect, forbidden in the vacuum.\footnote{%
In the vacuum the partial width for the decay $\omega \to \pi ^{+}\pi ^{-}$
is only $\sim 0.2\mathrm{MeV}$, and is due to the small isospin breaking and
the resulting $\rho -\omega $ mixing. In this paper we are not concerned
with this negligible effect.} In this paper we show that the matter-induced
width for this process is large. For $\omega $ moving with respect to the
medium with a momentum above $\sim 200\mathrm{MeV}$ (such momenta are easily
accessible in heavy-ion collisions) the corresponding width, at the nuclear
saturation density, is of the order of 100MeV. In addition, we find very
different behavior of the longitudinally and transversely polarized $\omega $
mesons, with the former ones being much wider than the latter ones.

Our calculation is made in the framework of an effective hadronic theory.
Mesons interact with the nucleons and $\Delta $ isobars, and the
interactions are assumed to have the usual form used in many other
calculations and fits. We work to the leading order in the nuclear density.
Based on the experience of other in-medium calculations we hope that this
approximation should be sufficient up to densities of the order of the
nuclear saturation density. To this leading order only the diagrams shown in
Fig. 1 contribute. In these diagrams the nucleon lines include the
propagation of occupied states of the Fermi sea. The ``bubble'' diagram (a)
has been analyzed by Wolf, Friman, and Soyeur in Ref. \cite{wolf}, where the
role of the $\omega -\sigma $ mixing mechanism has been pointed out. In this
process the $\omega $ meson is first converted, via interaction with the
nucleons, into the scalar-isoscalar $\sigma $ meson, which in turn decays
into two pions. The relevance of ``triangle'' diagrams (b) has been shown
out in Ref. \cite{bfh}. Note that in any formal counting scheme (low
density, chiral limit, large number of colors) the diagrams (a) and (b) are
of the same order and consistency requires to include both.
Our present calculation of $\omega \to \pi \pi $
in nuclear matter includes a further contribution of diagrams (c-d) with the 
$\Delta $(1232) isobar. Among other possible resonances, the $\Delta $ is
the most important one due to the large value of the $\pi N\Delta $ vertex
and small $\Delta -N$ mass splitting. 
\begin{figure}[tb]
\centerline{\psfig
{%
figure=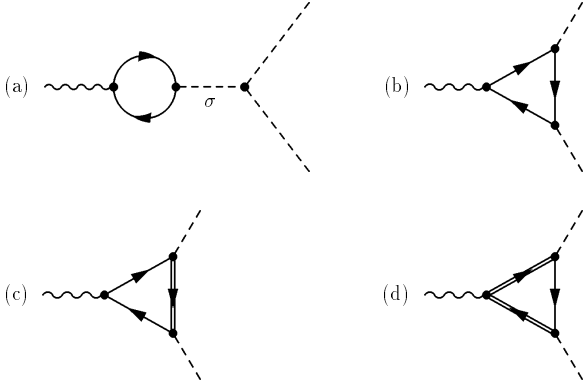,height=6.2cm,bbllx=75bp,bblly=315bp,bburx=422bp,bbury=573bp,clip=%
}} \label{ps}
\caption{Diagrams contributing to the $\omega \to \pi \pi $ amplitude in
nuclear medium. The incoming $\omega $ has momentum $q$ and polarization $%
\epsilon $. The outgoing pions have momenta $p$ and $q-p$. Diagrams (b-d)
have corresponding crossed diagrams, not displayed. The Feynman rules are
given in the text.}
\end{figure}

The solid line in Fig. 1 denotes the in-medium nucleon propagator, which can
be conveniently decomposed in the free and density parts \cite{chin}
\begin{eqnarray}
&&G(k) \equiv G_F(k)+G_D(k)\\
&&=(\not \!\!{k}+M)\left[\frac
1{k^2-M^2+i\varepsilon }+\frac{i\pi }{E_k}\delta (k_0-E_k)\theta
(k_F-|\mathbf{k}|)\right],\nonumber  \label{Nprop}
\end{eqnarray}
where $k$ is the nucleon four-momentum, $M$ denotes the nucleon mass, $E_k=%
\sqrt{M^2+\mathbf{k}^2}$, and $k_F$ is the Fermi momentum. The diagram (a)
is non-zero only when one propagator is $G_D$, and the other one $G_F$. The
only non-vanishing contributions in diagram (b) involve one $G_D$ propagator
and two $G_F$ propagators. Diagram (a) involves the intermediate $\sigma $%
-meson propagator, which we take in the form 
\begin{equation}
G_\sigma (k)={\frac 1{k^2-m_\sigma ^2+i\,m_\sigma \Gamma _\sigma -{\frac 14}%
\Gamma _\sigma ^2}}.  \label{sprop}
\end{equation}
Here the mass and the width of the $\sigma $ meson are chosen in such a way
that they reproduce effectively the experimental $\pi \pi $ scattering
length at $q^2=m_\omega ^2=(780\mathrm{MeV})^2$, which is the relevant
kinematic point for the process at hand. From this fit we find $m_\sigma
=789 $MeV and $\Gamma _\sigma =237$MeV. Note that $m_\omega $ and $m_\sigma $
are very close to each other, which enhances the amplitude obtained from
diagram (a) \cite{wolf}.

The double line in diagrams (c-d) denotes the $\Delta $ propagator 
\begin{eqnarray}
&&G_\Delta ^{\alpha \beta }(k)={\frac{\not \!\!{k}+M_\Delta }{k^2-M_\Delta
^2+i\,M_\Delta \Gamma _\Delta -{\frac 14}\Gamma _\Delta^2}} \\
&&\times \left[ -g^{\alpha
\beta }+{\frac 13}\gamma ^\alpha \gamma ^\beta +{\frac{2k^\alpha k^\beta }{%
3M_\Delta ^2}}+{\frac{\gamma ^\alpha k^\beta -\gamma ^\alpha k^\beta }{%
3M_\Delta }}\right] \nonumber.  \label{Dprop}
\end{eqnarray}
This formula corresponds to the usual Rarita-Schwinger definition \cite
{rarita,mukho} with the denominator modified in order to account for the
finite width of the $\Delta $ resonance, $\Gamma _\Delta =120$MeV.

We assume that the $\omega NN$ and $\omega \Delta \Delta $ vertices have the
form which follows from the minimum-substitution prescription and
vector-meson dominance applied to the nucleon and the Rarita-Schwinger
\cite{rarita} Lagrangians:
\begin{eqnarray}
&&V_{\omega NN}^\mu =g_\omega \gamma ^\mu ,\\
&&V_{\omega \Delta
\Delta }^{\mu \alpha \beta }=g_\omega \left[ -\gamma ^\mu g^{\alpha \beta
}+g^{\alpha \mu }\gamma ^\beta +g^{\beta \mu }\gamma ^\alpha +\gamma ^\alpha
\gamma ^\mu \gamma ^\beta \right] \nonumber.  \label{omv}
\end{eqnarray}
Possible anomalous couplings can be incorporated at the expense of having
more parameters. The results presented below do not depend qualitatively on
the form of the coupling, as long as it remains strong. The coupling
constant $g_\omega $ can be estimated from the vector dominance model. We
use $g_\omega =9$. For the $\pi NN$
vertex we use the pseudoscalar coupling, with the coupling constant $g_{\pi
NN}=$ $12.7$. The same value is used for $g_{\sigma NN}$. The $\sigma \pi
\pi $ coupling constant is taken to be equal to $g_{\sigma \pi \pi
}=12.8\,m_\pi $, where $m_\pi =139.6\mathrm{MeV}$ is the physical pion mass
(this value follows from the fit done to $\pi \pi $ scattering phase shifts
done in Ref. \cite{wolf}). The $\pi N\Delta $ vertex has the form
$V_{\pi N\Delta }^\mu =({f_{\pi N\Delta }}/{m_\pi})p^\mu {\vec T}$,
where $p^\mu $ is the pion momentum, $\vec T$ is the ${\frac 12}\rightarrow {%
\frac 32}$ isospin transition matrix, and the coupling constant $%
f_{\pi N\Delta }=2.1$ \cite{durso}.\footnote{%
There is another possible structure in the $\pi N\Delta$ coupling, of the
form $a\! \not \!\!{p} \gamma^\mu$. Our vertex corresponds to the
popular choice of the off-shell parameter $a$ set to zero.}

The amplitude, evaluated according to the diagrams depicted in Fig. 1 (a-d)
can be uniquely decomposed in the following Lorentz-invariant way: 
\begin{eqnarray}
\mathcal{M}=\epsilon ^\mu (Ap_\mu +Bu_\mu +Cq_\mu ),  \label{decomp}
\end{eqnarray}
where $p$ is the four-momentum of one of the pions, $q$ is the four-momentum
of the $\omega $ meson, $u$ is the four-velocity of nuclear matter, and $%
\epsilon $ specifies the polarization of $\omega $. Our calculation is
performed in the rest frame of nuclear matter, where $u=(1,0,0,0)$. In this
reference frame the amplitude $\mathcal{M}$ vanishes for vanishing
3-momentum $\mathbf{q}$, as requested by rotational invariance. Hence, the
process $\omega \rightarrow \pi \pi $ occurs only when the $\omega $ moves
with respect to the medium.

The expression for the decay width reads
\begin{eqnarray}
&&\Gamma=\frac 123\frac 1{n_s}\sum_s\frac 1{2q_0}\int \frac{d^3p}{(2\pi )^32p_0%
}\int \frac{d^3p^{\prime }}{(2\pi )^32p_0^{\prime }}
\nonumber \\ &&\times |\mathcal{M}|^2(2\pi
)^4\delta ^{(4)}(q-p-p^{\prime }),  \label{widthg}
\end{eqnarray}
where the factor $\frac 12$ is the symmetry factor when the decay proceeds
into two neutral pions, the factor of $3$ accounts for the isospin
degeneracy of the final pion states (\emph{i.e.} neutral and charged pions), 
$n_s$ is the number of spin states of the $\omega $ meson, and $\sum_s$
denotes the sum over these spin states ($q$ , $p$ and $p^{\prime }=q-p$ are
the four-momenta of the $\omega $ meson, and the two pions, respectively).

\begin{figure}[t]
\vspace{0mm} \epsfxsize = 7.5 cm \centerline{\epsfbox{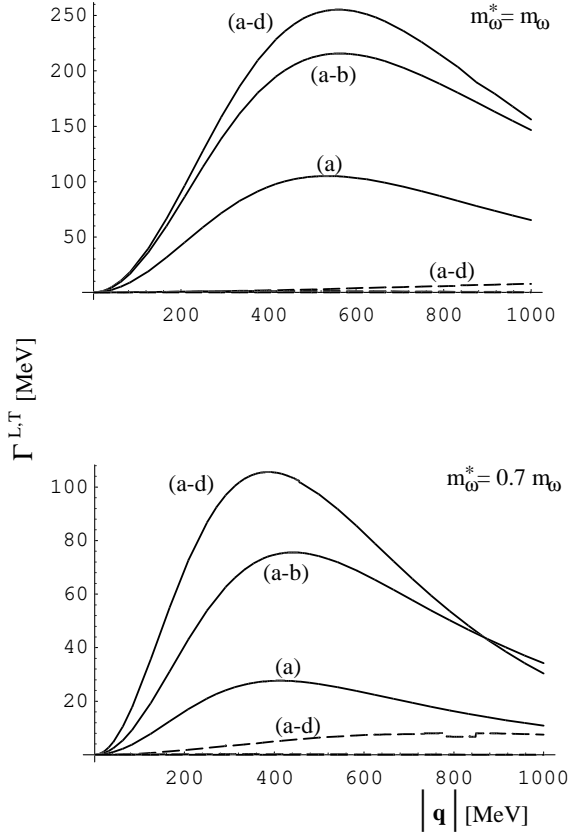}} \vspace{0mm}
\label{diag}
\caption{The in-medium width of the $\omega $ meson plotted as a function of
its 3-momentum $|\mathbf{q|}$. The solid (dashed) lines correspond to the
longitudinal (transverse) mode. The labels (a), (a-b) and (a-d) refer to
Fig. 1. They indicate the diagrams included in the calculation: (a) $%
\omega-\sigma$ mixing mechanism, (a-b) $\omega-\sigma$ mixing together with
the ``triangle'' nucleon diagrams, (a-d) the full result, including the
contribution from the $\Delta$ isobar.}
\end{figure}

We take the effort to analyze separately the longitudinally and transversely
polarized $\omega $, since the presence of the medium results in different
behavior of these states. Transversely polarized $\omega $ has two helicity
states ($n_s=2$), with projections $s=\pm 1$ on the direction of $\mathbf{q}$%
, and the longitudinally polarized $\omega $ has one helicity state ($n_s=1$%
), with the corresponding projection $s=0$. An explicit calculation yields
\begin{eqnarray}
&&\sum_{s=\pm 1}\varepsilon _{(s)}^\mu \varepsilon _{(s)}^{*\nu }
\equiv -T^{\mu \nu }= \nonumber \\
&&=\frac{%
(q^\mu -q\cdot u\ u^\mu )(q^\nu -q\cdot u\ u^\nu )}{q\cdot q-(q\cdot u)^2}%
-g^{\mu \nu }+u^\mu u^\nu ,  \nonumber \\
&&\varepsilon _{(s=0)}^\mu \varepsilon _{(s=0)}^{*\nu }
\equiv -L^{\mu \nu } = \nonumber \\
&&=-\frac{(q^\mu
-q\cdot u\ u^\mu )(q^\nu -q\cdot u\ u^\nu )}{q\cdot q-(q\cdot u)^2}+\frac{%
q^\mu q^\nu }{q\cdot q}-u^\mu u^\nu .  \label{tensors}
\end{eqnarray}
The tensors $T^{\mu \nu }$ and $L^{\mu \nu }$ are defined in
such a way that they are projection operators. In addition, $T^{\mu \nu
}q_\nu =0$ and $L^{\mu \nu }q_\nu =0$, which reflects current conservation,
and $T^{\mu \nu }u_\nu =0$. From relations (\ref{decomp}) and (\ref{tensors}%
) in Eq. (\ref{widthg}) we find that
\begin{eqnarray}
&&|\mathcal{M}_T|^2=|A|^2p_\mu (-T^{\mu \nu })p_\nu ,\\
&&|\mathcal{M}%
_L|^2=(A^{*}p_\mu +B^{*}u_\mu )(-L^{\mu \nu })(Ap_\nu +Bu_\nu ). \nonumber
\label{widthL}
\end{eqnarray}
Note that the value of the coefficient $C$ is irrelevant for our
calculation. For the diagram (a) we find $A=0,$ $B\neq 0,$ hence this
diagram contributes only to the width of the longitudinal mode. Diagrams
(b-d) have $A\neq 0,$ $B\neq 0,$ and contribute to the width of both the
longitudinal and transverse modes.

As we have said, we evaluate the amplitude $\mathcal{M}$ to leading order in
the baryon density. This leads to a simplification. The integrals of the
form $\int_0^{k_F}k^2dk\,f(k)$ arising in our calculation are replaced by $%
f(0)\int_0^{k_F}k^2dk$, which is proportional to baryon density, $\rho _B$.
Consequently, the widths $\Gamma ^{L,T}\sim $ $\rho _B^2$.

In Fig. 2 we present our numerical results at the nuclear saturation
density, $\rho _B=0.17\mathrm{fm}^{-3}$. We show $\Gamma ^L$ (solid lines) 
and $%
\Gamma ^T$ (dashed lines) plotted as functions of $|\mathbf{q}|$. The upper
part of the plot is for $m_\omega ^{*}=m_\omega =780\mathrm{MeV}$, i.e. the
value of the $\omega $ mass is not modified by the medium. The lower part is
for $m_\omega ^{*}=0.7m_\omega $. In both cases we reduce the value of the
in-medium nucleon mass to 70 \% of its vacuum value, $M^{*}=0.7M,$ which is
a typical number at the nuclear saturation density. We also reduce by the
same factor the mass of the $\Delta $, \emph{i.e.} $M_\Delta
^{*}=0.7M_\Delta $, since it is expected to behave similarly to the nucleon.
The labels indicate which diagrams of Fig. 1 have been included. The
complete result corresponds to the case (a-d). The case (a) reproduces the
result of Ref. \cite{wolf}.
We note that the inclusion of subsequent processes of Fig. 1
substantially increases the result. All the curves start as $\mathbf{q}^2$
at low $|\mathbf{q}|$. The longitudinal width reaches a maximum at $|\mathbf{%
q}|\sim $ a few hundred MeV, and the value at the peak is large: 250MeV for $%
m_\omega ^{*}=m_\omega $ and 100MeV for $m_\omega ^{*}=0.7m_\omega $. The
transverse width is strictly zero with diagram (a), less than 1MeV with
diagrams (a-b), reach a few MeV when the diagrams with the $\Delta $ are
included. Qualitatively similar results follow for other choices of
parameters. One should bare in mind that the effect is proportional to $\rho
_B^2$, hence may be much larger at higher densities.

Our main conclusions are:
1)  nuclear matter induces the $\omega \to \pi \pi $ transitions with
large partial widths,
2)  the widths depend strongly on the three momentum of the $\omega $
with respect to the medium, $|\mathbf{q}|$,
3) the longitudinal mode is much wider than the transverse mode.

The results obtained mean that in a hadron gas, such as created in a
heavy-ion collision, the propagation of longitudinally polarized $\omega $
meson is inhibited when the momentum $|\mathbf{q}|$ is nonzero. This will
cause a depletion in the population of the $\omega $ mesons. Such effects
should be included in Monte-Carlo simulations of heavy-ion collisions.

The important question is to what extent can the discussed process influence
shape of the dilepton-production cross-sections in relativistic heavy-ion
collisions. Our results can be used to calculate the cross section for the $%
\pi \pi $ annihilation into dileptons occurring in the $\omega $ channel.
This mechanism has been analyzed for the first time in Ref. \cite{wolf}. The
calculation of the annihilation cross section requires the knowledge of the
same amplitude that has been used in the calculation of the omega width, see
Fig. 3. In Ref. \cite{wolf} only the $\omega -\sigma $ mixing term was
included in this amplitude (diagram (a) of Fig. 1). In our present
calculation we take into account additional diagrams shown in Figs. 1 (b) -
(d). Moreover, we take into consideration differences in the propagation of
the transverse and longitudinal modes, which is important due to the large
difference observed in the widths. 
\begin{figure}[tb]
\centerline{\psfig
{%
figure=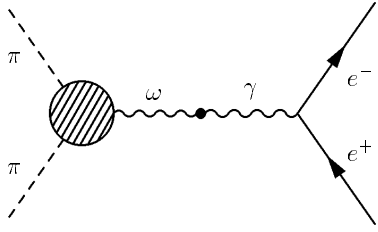,height=3.4cm,bbllx=230bp,bblly=390bp,bburx=408bp,bbury=505bp,clip=%
}} \label{annih}
\caption{Pion annihilation process in the $\omega $ channel. The blob
corresponds to the amplitude shown in Fig. 1, and the dot denotes the
vector-meson-dominance conversion factor $e m_\omega^2/(2g_\omega )$.}
\end{figure}
The $\pi \pi $ annihilation cross section corresponding to Fig. 3, averaged
over the incoming pion momenta at fixed total four-momentum $q=(q^0,\mathbf{q%
})$, can be written in the compact form
\begin{equation}
\sigma ={\frac{8\pi ^2q^0m_\omega ^4}{%
9q^2(q^2/4-m_\pi ^2)}}\left( {\frac \alpha {g_\omega }}\right) ^2\left(
2|G_\omega ^T|^2\Gamma ^T+|G_\omega ^L|^2\Gamma ^L\right) ,  \label{sigma}
\end{equation}
where
\begin{equation}
|G_\omega ^{T,L}|^2={\frac 1{[q^2-m_\omega ^2-
\frac{1}{4}(\Gamma_0+\Gamma^{T,L})^2]^2+m_\omega
^2(\Gamma _0+\Gamma ^{T,L})^2}}  \label{omprop2}
\end{equation}
is the modulus of the $\omega $ propagator squared. The widths $\Gamma ^T$
and $\Gamma ^L$ should be calculated in the way described above with the
only difference that the physical mass of $\omega $ is now replaced by the
invariant dilepton mass $\sqrt{q^2}$. The quantity $\Gamma _0$ denotes the width of
the $\omega $ at vanishing three momentum, which is
due to other effects, such as $%
\omega \to \pi \pi \pi $. With $\Gamma _0\sim 10\mathrm{MeV}$ \cite{klingl}
and our values for $\Gamma ^{T,L}$ we obtain the cross section from Eq. (\ref
{sigma}), which is typically a fraction of a microbarn. This is to be compared to
$3.5\mu{\rm b}$ from the decay via the $\rho $ resonance \cite{wolf}. Note
that large widths $\Gamma^{T,L}$ in Eq. (\ref{sigma}) do not
increase $\sigma$,
since they also appear
in the denominator of Eq. (\ref{omprop2}).
In fact, there are optimum widths $\Gamma ^{T,L}\sim
\Gamma_0$ at which the cross section is the largest.
A further increase of the
widths $\Gamma ^{T,L}$ decreases the cross section.
At the point $q^2=m_\omega ^2$ and
with our numbers from Fig. 2 we find that the contribution of the
longitudinal modes to Eq. (\ref{sigma}) is negligible, while the contribution
from the transverse modes at $|\mathbf{q}|=400\mathrm{MeV}$ equals
$0.4\mu{\rm b}$ for $m_\omega^*=m_\omega$ (where $\Gamma^T=1.8{\rm MeV}$),
and $1.4\mu{\rm b}$ for $m_\omega^*=0.7m_\omega$
(where $\Gamma^T=5{\rm MeV}$).
We stress that the
numbers quoted above are almost entirely due to the diagrams
with the $\Delta$. Without the processes (b-d) of Fig. 1 the dilepton
production via mechanism of Fig. 3 would be about a factor of 10 smaller.
In conclusion, the process of Fig. 3 may be significant for the
dilepton production in heavy-ion collisions.

If the dilepton-production
experiments measured the three-momentum $\mathbf{q}$ of
the dilepton pair coming from a vector-meson decay, then they should observe
different behavior at different values of $|\mathbf{q|}$. Such measurement
would be very helpful for a better understanding of meson dynamics in the
nuclear medium.

Our last remark refers to the final state interactions, which can be
important \cite{durso}. The pions emitted in processes of
Fig. 1 can interact in the final channel. This will result in an appropriate
modification the $\omega \to \pi\pi$ amplitude. The full analysis of the
final-state interactions requires a model for the $\pi \pi$ scattering
amplitude, as well as solving a Lipmann-Schwinger equation. This is beyond
the scope of this paper. Note, however, that the diagram (a), which includes
the intermediate $\sigma$ state, does in fact account for final-state
interactions. In this process the pions form a resonance in the $S$-channel,
which enhances the amplitude. Similar rescattering can also occur for the
diagrams (b-d). Thus, the final-state interactions are only partially
included in our analysis.

\begin{acknowledgement}
We thank Bengt Friman for numerous valuable comments and for the suggestion
to include the $\Delta$ in the presented analysis.
\end{acknowledgement}


\end{document}